\documentclass[epj]{svjour}
\usepackage{amsmath}  % pour aligner les équations
\usepackage{amssymb}  % symboles mathématiques
\usepackage{graphics}
\begin{document}
\title{Study of the shape coexistence in the $^{96}$Zr, $^{96}$Mo, $^{96}$Ru isobars}
\author{R. Budaca\inst{1,2}, P. Buganu\inst{1}, F. El Ouardi\inst{3} \and A. Lahbas\inst{3,4}% etc
}                     % Do not remove
\institute{"Horia Hulubei"-National Institute for R\&D in Physics and Nuclear Engineering, Str. Reactorului 30, RO-077125, POB-MG6 Bucharest-M\v{a}gurele, Romania \and Academy of Romanian Scientists, Splaiul Independentei St., No. 54, Bucharest, P. O. 050094, Romania \and ESMaR, Department of Physics, Faculty of Science, Mohammed V University in Rabat, Rabat 10000, Morocco \and High Energy Physics and Astrophysics Laboratory, Department of Physics, Faculty of Science Semlalia, Cadi Ayyad University, P. O. B. 2390, Marrakesh 40000, Morocco}

\date{Received: date / Revised version: date}
\abstract{
Three stable isobars, $^{96}_{40}$Zr$_{56}$, $^{96}_{42}$Mo$_{54}$ and $^{96}_{44}$Ru$_{52}$, which are in the vicinity of the harmonic oscillator proton shell closure $Z=40$ and the spin-orbit neutron shell closure $N=50$, are investigated for the presence of the shape coexistence and mixing phenomena. The ground state deformation of these isobars is extracted from the potential energy surface determined with the Covariant Density Functional Theory using a density-dependent point-coupling interaction, while the excited states are described involving the Bohr-Mottelson Hamiltonian with octic potential for both axially symmetric and $\gamma$-unstable quadrupole deformations. Within the broader view of the two approaches, the obtained results clearly highlight the significant contribution of these phenomena to the structure of the states of these nuclei.
%
%\PACS{
 %     {PACS-key}{discribing text of that key}   \and
 %     {PACS-key}{discribing text of that key}
  %   } % end of PACS codes
} %end of abstract
\maketitle
\section{Introduction}
The shape coexistence phenomenon \cite{Morinaga}, which has been widely studied over the years \cite{Heyde1}, seems to be often invoked lately in the interpretation of the structure of the low-lying quadrupole states \cite{Wood1,Garrett,Bonatsos1}. This is due to the fact that the energy levels, as well as the electromagnetic transitions between these states, are better reproduced if the presence of this phenomenon is considered in some nuclei. Therefore, certain signatures have been proposed to identify the nuclei suspected of manifesting such behavior \cite{Bonatsos2}. Most often, the first excited $0^{+}$ state is taken as reference when a candidate nucleus is selected. If the energy of this state is very low, respectively the monopole transition from this excited $0^{+}$ state and the ground state has an appreciable value, then the respective nucleus could exhibit shape coexistence \cite{Wood2}. Moreover, the idea of the existence of islands of shape coexistence has been quite recently advanced \cite{Martinou1,Martinou2}. This is very useful in numerical applications of the models to the experimental data, simplifying the selection procedure of the candidate nuclei. Concerning the theoretical models involved in studying the shape coexistence, there is a whole variety depending on the adopted approach, but also on the studied physical quantities related to the shape coexistence phenomenon. We give here some examples without claiming to cover all of them, this goal being beyond the purpose of the present study. Thus, one has models based on covariant density functional theory \cite{Vretenar2005,Niksic2011,Yang}, shell structure \cite{Mayer,Poves}, proxy-SU(3) \cite{Martinou1,Martinou2,Bonatsos3}, interacting boson approximation \cite{Iachello,Nomura,Maya}, partial dynamical symmetry \cite{Leviatan}, ab initio \cite{Zhou,Hu}, a five-dimensional collective Hamiltonian based on covariant density functional theory \cite{Majola,Matsuyanagi}, general collective approach \cite{Gneuss,Hess,Troltenier}, algebraic collective approach \cite{Rowe,Welsh,Georgoudis}, Bohr-Mottelson Hamiltonian with polynomial potentials \cite{Bohr1,Bohr2,Budaca1,Buganu1} and others \cite{Heyde1,Wood1,Bonatsos1,BudacaAI}, each of them offering a different perspective in looking to the shape coexistence phenomenon, but more importantly a better understanding of it.

Two alternative descriptions, one microscopic and the other phenomenological, are complementarily employed in the present study for a more accurate view of the shape properties in these nuclei. The former is the Covariant Density Functional Theory (CDFT) with a Density- Dependent Point-Coupling X (DD-PCX) parametrization  \cite{Vretenar2005,Niksic2011,PCX_Optimization,Ring2001,Tian2009}, which is used to extract information about the ground state deformation, while the latter is represented by the Bohr-Mottelson Hamiltonian \cite{Bohr1,Bohr2} with an Octic Potential (BMH-OP) for $\gamma$-unstable \cite{Buganu1} and prolate \cite{Budaca2} deformations applied to investigate the properties of the excited states. In this case, the octic potential allows a description of shape coexistence between an approximately spherical shape and a well-deformed one, prolate or $\gamma$-unstable. Depending on the height of the barrier, which separates the two minima, one can have shape coexistence with and without a mixing of the shapes \cite{Buganu1,Budaca3}. This ability of the model is related to the numerical solvability method applied for the Hamiltonian, being thus a very general solution compared to quasi-exact solvable methods \cite{Ushveridze}. The latter method applied for the sextic potential offers exact solutions, but only to a part of the eigenvalue problem \cite{Levai1}. Nevertheless, this method proved to be very suitable to describe shape phase transitions and their critical points \cite{Lahbas,Levai2}. In turn, the numerical method using as a basis the Bessel functions of the first kind \cite{Taseli}, allowed a description of the shape coexistence and mixing phenomena \cite{Budaca1,Budaca2,Budaca3,Budaca4,Budaca5,Mennana1,Mennana2,Benjedi1,Buganu2,Buganu3,Benjedi2}. Recently, the method has been extended to the octic potential \cite{Buganu1,Budaca2} increasing in this way the range of applications to the experimental data. Preliminary applications with the octic potential have been done for the $^{98-106}$Ru and $^{106-116}$Cd even-even isotopic chains \cite{Buganu1,Budaca2}, while new ones are done  in the present study for the $^{96}$Zr, $^{96}$Mo and $^{96}$Ru isobars.

The plan of the present work is the following. The two models, CDFT with DD-PCX and BMH-OP, are introduced in Section 2, while the results for the $^{96}$Zr, $^{96}$Mo and $^{96}$Ru isobars are presented and discussed in Section 3. Finally, the main achievements of the study are highlighted in Section 4.

\section{Description of the models}
Two models are involved in the present study, a microscopic and a phenomenological one, respectively. These models are briefly introduced in the following two subsections.

\subsection{Covariant Density Functional Theory with Density-Dependent Point-Coupling X}
The nuclear ground-state properties investigated in this work are described within
the framework of Covariant Density Functional Theory (CDFT), implemented through
the relativistic Hartree--Bogoliubov (RHB) approach with the density-dependent
point-coupling interaction DD-PCX \cite{Vretenar2005,Niksic2011,PCX_Optimization}.
Within this framework, nucleons are treated as Dirac particles evolving in
self-consistent mean fields arising from zero-range four-fermion contact interactions,
which guarantee Lorentz covariance and a proper treatment of spin degrees of
freedom \cite{Niksic2011,Ring2001}. The effective Lagrangian density of the
density-dependent point-coupling model takes the form \cite{PCX_Optimization}:
\begin{equation}
\begin{aligned}
\mathcal{L} ={}& \bar{\psi}(i\gamma^\mu \partial_\mu - m)\psi  - \frac{1}{2}\alpha_S(\rho)(\bar{\psi}\psi)^2 \\
& - \frac{1}{2}\alpha_V(\rho)(\bar{\psi}\gamma^\mu\psi)(\bar{\psi}\gamma_\mu\psi) \\
& - \frac{1}{2}\alpha_{TV}(\rho)(\bar{\psi}\vec{\tau}\gamma^\mu\psi)(\bar{\psi}\vec{\tau}\gamma_\mu\psi) \\
& - \frac{1}{2}\delta_S(\partial_\nu \bar{\psi}\psi)(\partial^\nu \bar{\psi}\psi)  - e\bar{\psi}\gamma^\mu A_\mu \frac{1-\tau_3}{2}\psi,
\end{aligned}
\end{equation}
where $\psi$ represents the nucleon Dirac field with mass $m$. The derivative
coupling term proportional to $\delta_S$ accounts for finite-range effects,
which are essential for a correct description of nuclear surface
properties \cite{Niksic2011}. The density dependence of the coupling functions
is expressed as \cite{PCX_Optimization}:
\begin{equation}
\alpha_i(\rho) = a_i + (b_i + c_i x)e^{-d_i x}, \qquad (i = S, V, TV),
\end{equation}
where $x = \rho/\rho_{\mathrm{sat}}$ and $\rho_{\mathrm{sat}}$ is the saturation
density of symmetric nuclear matter. The labels S, V, and TV correspond to the
isoscalar-scalar, isoscalar-vector, and isovector-vector channels of the nucleon
interaction. The DD-PCX parameterization involves ten coupling constants governing
these channels, which are listed in Table~\ref{tab:DDPCX}.

Pairing correlations in open-shell nuclei are handled self-consistently within
the RHB framework \cite{Vretenar2005,Niksic2011}. The total energy density
functional is written as:
\begin{equation}
E_{\mathrm{RHB}}[\rho,\kappa] = E_{\mathrm{RMF}}[\rho] + E_{\mathrm{pair}}[\kappa],
\end{equation}
where $\rho$ denotes the normal density matrix and $\kappa$ is the pairing tensor.
The pairing energy contribution reads \cite{Vretenar2005}:
\begin{equation}
E_{\mathrm{pair}}[\kappa] =
\frac{1}{4}
\sum_{n_1 n'_1}
\sum_{n_2 n'_2}
\kappa^{*}_{n_1 n'_1}
\langle n_1 n'_1 | V^{PP} | n_2 n'_2 \rangle
\kappa_{n_2 n'_2},
\end{equation}
where $V^{PP}$ is the pairing interaction and the indices refer to the Dirac
spinor quantum numbers. The pairing force is adopted in a separable form in
momentum space, which in coordinate space is expressed as \cite{Tian2009}:
\begin{equation}
V^{PP}(\mathbf{r}_1,\mathbf{r}_2,\mathbf{r}'_1,\mathbf{r}'_2)
= -G\,\delta(\mathbf{R}-\mathbf{R}')
P(\mathbf{r})P(\mathbf{r}'),
\end{equation}
with the center-of-mass and relative coordinates defined as
\begin{equation}
\mathbf{R} = \frac{1}{\sqrt{2}}(\mathbf{r}_1+\mathbf{r}_2),
\qquad
\mathbf{r} = \frac{1}{\sqrt{2}}(\mathbf{r}_1-\mathbf{r}_2),
\end{equation}
and a Gaussian form factor of the form
\begin{equation}
P(\mathbf{r}) =
\frac{1}{(4\pi a^2)^{3/2}}
\exp\left(-\frac{r^2}{2a^2}\right).
\end{equation}
This choice of pairing interaction preserves translational invariance,
being shown to reliably describe pairing properties throughout the nuclear
chart \cite{Tian2009}. Finally, the potential energy surfaces are computed
through constrained RHB calculations employing the method of quadratic
constraints \cite{Niksic2011}. The augmented energy functional to be minimized is:
\begin{equation}
E' = \langle \hat{H}_{tot} \rangle
+ \sum_{\mu=0,2}
C_{2\mu}
\left(\langle \hat{Q}_{2\mu} \rangle - q_{2\mu}\right)^2,
\end{equation}
where $\langle \hat{H}_{tot} \rangle$ is the expectation value of the total
Hamiltonian, $\hat{Q}_{20} = 2z^2 - x^2 - y^2$ and $\hat{Q}_{22} = x^2 - y^2$
are the mass quadrupole operators, $q_{2\mu}$ are the target constraint values,
and $C_{2\mu}$ are the corresponding stiffness constants. This approach enables
a systematic mapping of the potential energy surface as a function of the
quadrupole deformation parameters $(\beta_2, \gamma)$.

\begin{table}[htbp]
\centering
\caption{Parameters of the density-dependent point-coupling interaction (DD-PCX)
\cite{PCX_Optimization}.
The saturation density is $\rho_{\mathrm{sat}} = 0.152~\mathrm{fm}^{-3}$ and the nucleon
mass is $m = 939~\mathrm{MeV}$.}
\label{tab:DDPCX}
\begin{tabular}{ccc}
\hline
Parameter & Value & Unit \\
\hline
$a_S$ & $-10.979243836$ & fm$^{2}$ \\
$b_S$ & $-9.038250910$  & fm$^{2}$ \\
$c_S$ & $-5.313008820$  & fm$^{2}$ \\
$d_S$ & $1.379087070$   & -- \\
$a_V$ & $6.430144908$   & fm$^{2}$ \\
$b_V$ & $8.870626019$   & fm$^{2}$ \\
$d_V$ & $0.655310525$   & -- \\
$b_{TV}$ & $2.963206854$ & fm$^{2}$ \\
$d_{TV}$ & $1.309801417$ & -- \\
$\delta_S$ & $-0.878850922$ & fm$^{4}$ \\
\hline
$G_n$ & $-800.663126037$ & MeV\,fm$^{3}$ \\
$G_p$ & $-773.776776597$ & MeV\,fm$^{3}$ \\
\hline
\end{tabular}
\end{table}

\subsection{Bohr-Mottelson Hamiltonian with octic potential}
The phenomenological Hamiltonian has the expression introduced in \cite{Bohr1,Bohr2}:
\begin{eqnarray}
\hat{H}&=&-\frac{\hbar^{2}}{2B}\left[\frac{1}{\beta^{4}}\frac{\partial}{\partial\beta}\beta^{4}\frac{\partial}{\partial\beta}+\frac{1}{\beta^{2}\sin3\gamma}\frac{\partial}{\partial\gamma}\sin3\gamma\frac{\partial}{\partial\gamma}\right]\nonumber\\
&&+\frac{\hbar^{2}}{8B\beta^{2}}\sum_{k=1}^{3}\frac{\hat{L}_{k}^{2}}{\sin^{2}\left(\gamma-\frac{2\pi}{3}k\right)}+V(\beta,\gamma),
\label{Hamiltonian}
\end{eqnarray}
with the following notations introduced for: $B-$the mass parameter, $\hbar-$the reduced Planck constant, $\hat{L}_{k}-$the angular momentum projections in the intrinsic reference frame, $\beta$ and $\gamma-$the intrinsic deformation coordinates, and $V(\beta,\gamma)$-the energy potential. A first step toward the separation of variables is to consider the energy potential of the form \cite{Fortunato}:
\begin{equation}
V(\beta,\gamma)=V_{1}(\beta)+\frac{1}{\beta^{2}}V_{2}(\gamma),
\label{potential}
\end{equation}
respectively, to choose a wave function of the following form:
\begin{equation}
\Psi(\beta,\gamma,\theta_{1},\theta_{2},\theta_{3})=F(\beta)\psi(\gamma,\theta_{1},\theta_{2},\theta_{3}).
\end{equation}
Here, $\theta_{1}$, $\theta_{2}$ and $\theta_{3}$ are the Euler angles associated with rotations of the nucleus. This selection leads to an exact separation of the $\beta$ degree of freedom from the other four variables:
\begin{equation}
\left[-\frac{1}{\beta^{4}}\frac{\partial}{\partial\beta}\beta^{4}\frac{\partial}{\partial\beta}+v_{1}(\beta)-\varepsilon+\frac{\Lambda}{\beta^{2}}\right]F(\beta)=0,
\label{eqbeta}
\end{equation}
\begin{eqnarray}
&&\left[-\frac{1}{\sin3\gamma}\frac{\partial}{\partial\gamma}\sin3\gamma\frac{\partial}{\partial\gamma}+\frac{1}{4}\sum_{k=1}^{3}\frac{\hat{L}_{k}^{2}}{\sin^{2}\left(\gamma-\frac{2\pi}{3}k\right)}\right]\\
&&\times\psi(\gamma,\theta_{i})+v_{2}(\gamma)\psi(\gamma,\theta_{i})=\Lambda\psi(\gamma,\theta_{i}),\;i=1,2,3.\nonumber
\label{eqgamma}
\end{eqnarray}
Here, by $\Lambda$ is denoted the separation constant, while $\varepsilon=\frac{2B}{\hbar^{2}}E$, $v_{1}(\beta)=\frac{2B}{\hbar^{2}}V_{1}(\beta)$ and $v_{2}(\gamma)=\frac{2B}{\hbar^{2}}V_{2}(\gamma)$ are the reduced energy and potentials, respectively. Eq. (\ref{eqbeta}) has been numerically solved in \cite{Buganu1,Budaca2} for an octic potential:
\begin{equation}
v_{1}(\beta)=\beta^{2}+b_{1}\beta^{4}+b_{2}\beta^{6}+b_{3}\beta^{8},
\label{OP}
\end{equation}
where $b_{1}$, $b_{2}$ and $b_{3}$  are free parameters. The functions corresponding to the octic potential are written in a basis of the Bessel functions of the first kind \cite{Budaca1}:
\begin{equation}
F_{\xi,\nu}(\beta)=\sum_{n=1}^{n_{Max}}A_{n}^{\xi}f_{n,\nu}(\beta),
\label{totwav}
\end{equation}
where,
\begin{equation}
f_{n,\nu}(\beta)=\frac{\sqrt{2}}{\beta_{w}}\frac{\beta^{-\frac{3}{2}}J_{n,\nu}\left(\frac{z_{n,\nu}}{\beta_{w}}\beta\right)}{J_{\nu+1}(z_{n,\nu})}
\end{equation}
are solutions of the equation in the $\beta$ variable for an infinite square well potential \cite{Iachello1,Iachello2}. Here, $\xi=n_{\beta}+1$ is related to the $\beta$-vibration quantum number $n_{\beta}=0,1,2,...$, $\nu$ is the index of the Bessel function $J_{n,\nu}$, $n_{Max}$ gives the dimension of the basis, $A_{n}^{\xi}$ are the eigenvector components, $z_{n,\nu}$ is the zero of the Bessel function, while $\beta_{w}$ is the position of the infinite square well potential. The general matrix element for this function is \cite{Buganu1,Budaca2}:
\begin{equation}
H_{mn}=\left(\frac{z_{n,\nu}}{\beta_{w}}\right)^{2}\delta_{mn}+\beta_{w}^{2}I_{mn}^{(\nu,1)}+\sum_{i=2}^{4}b_{i-1}\beta_{w}^{2i}I_{mn}^{(\nu,i)},
\label{matrixelements}
\end{equation}
where,
\begin{eqnarray}
I_{mn}^{(\nu,i)}&=&\frac{2}{J_{\nu+1}(z_{m,\nu})J_{\nu+1}(z_{n,\nu})}\times\\
&&\int_{0}^{1}J_{\nu}(z_{m,\nu}x)J_{\nu}(z_{n,\nu}x)x^{2i+1}dx, \;x=\beta/\beta_{w}.\nonumber
\end{eqnarray}
This matrix element is calculated for the index of the Bessel function given by:
\begin{equation}
\nu=\sqrt{\Lambda+\frac{9}{4}},
\end{equation}
where, $\Lambda$ is the eigenvalue of the $\gamma$ equation (\ref{eqgamma}). Two solutions are considered in the present study for Eq. (\ref{eqgamma}). The first one is for a $\gamma$-unstable system \cite{Buganu1} with
\begin{equation}
\nu=\sqrt{\tau(\tau+3)+\frac{9}{4}}=\tau+\frac{3}{2}, \;\tau=0,1,2,...,
\end{equation}
where $\tau$ is called the seniority quantum number \cite{Bohr1,Bohr2,Bes,Rakavy}. The total energy in this case is \cite{Buganu1}:
\begin{equation}
E_{\xi,\tau,L}=\frac{\hbar^{2}}{2B}\left[\varepsilon_{\xi,\tau}(b_{1},b_{2},b_{3})+cL(L+1)\right].
\label{eunstable}
\end{equation}
The last term, $L(L+1)$, is the eigenvalue of the $SO(3)$ symmetry operator $\hat{L}^{2}$ and it was added to broke the degeneracy over $\tau$ \cite{Caprio}. This term does not change in any way the quantum structure of the problem, because $\hat{L}^{2}$ commutes with $\hat{H}$.
The second solution is for an axially-symmetric deformation \cite{Budaca2}, with
\begin{equation}
\nu=\sqrt{\Lambda+\frac{9}{4}}=\sqrt{\frac{L(L+1)-K^2}{3}+6qn_{\gamma}+\frac{9}{4}},
\end{equation}
respectively the energies:
\begin{equation}
E_{\xi,n_{\gamma},L,K}=\frac{\hbar^{2}}{2B}\varepsilon_{\xi,n_{\gamma},L,K}(b_{1},b_{2},b_{3},q),
\end{equation}
Here, $n_{\gamma}$ is the $\gamma$-vibration quantum number, $K$ is the quantum number of the $\hat{L}_{3}$ angular momentum projection on the $z$-axis, while $q$ is the parameter of the potential in the $\gamma$ variable \cite{Budaca2,Budaca5}. Here, $n_{\gamma}=0$ for the ground and $\beta$ bands, respectively $n_{\gamma}=1$ for the first $\gamma$ band. In both cases, $\gamma$-unstable and prolate deformations, the $B(E2)$s electromagnetic transition probabilities are calculated with the transition operator \cite{Wilets}:
\begin{eqnarray}
T_{2,\mu}^{(E2)}&=&t\beta D_{\mu,2}^{(2)}(\theta_{i})\cos\gamma+\nonumber\\
&&\frac{t\beta}{\sqrt{2}}\left[D_{\mu,2}^{(2)}(\theta_{i})+D_{\mu,-2}^{(2)}(\theta_{i})\right]\sin\gamma,
\label{tranop}
\end{eqnarray}
where $t=3R^{2}Ze\beta_{M}/4\pi$ depends on the nuclear radius $R=R_{0}A^{1/3}$, charge number $Z$, elementary charge $e$ and a scaling factor $\beta_M=\beta_{2}/\beta$. Here, $R_{0}=1.2$ fm, $A$ is the mass number, $\beta_{2}$ the quadrupole deformation, while by $D(\theta_{i})$ denote the Wigner matrices \cite{Wigner}. For the $\gamma$-unstable case one has \cite{Buganu1}:
\begin{eqnarray}
&&B(E2;\xi\tau L\rightarrow\xi'\tau' L')=\left(\frac{3R^{2}Ze}{4\pi}\right)^{2}\beta_{M}^{2}\times\nonumber\\
&&(\tau'\alpha'L';112||\tau\alpha L)^{2}\langle\tau|||Q|||\tau'\rangle^{2}(B_{\xi\tau;\xi'\tau'})^{2},
\label{BE2}
\end{eqnarray}
where, $(\tau_{1}\alpha_{1}L_{1};\tau_{2}\alpha_{2}L_{2}||\tau_{3}\alpha_{3}L_{3})$ is the $SO(5)$ Clebsch-Gordon coefficient \cite{Rowe1}, while
\begin{equation}
\langle\tau|||Q|||\tau'\rangle=\sqrt{\frac{\tau}{2\tau+3}}\delta_{\tau,\tau'+1}+\sqrt{\frac{\tau+3}{2\tau+3}}\delta_{\tau,\tau'-1},
\label{tauop}
\end{equation}
is the $SO(5)$ reduced matrix element of the quadrupole moment \cite{Rowe2,Rowe3}.
The integral over $\beta$ is denoted here by:
\begin{equation}
B_{\xi\tau;\xi'\tau'}=\langle F_{\xi,\tau}(\beta)|\beta|F_{\xi',\tau'}(\beta)\rangle.
\end{equation}
In turn, the $B(E2)$s for the axially-symmetry are given by \cite{Budaca2}:
\begin{eqnarray}
&&B(E2;\xi,n_{\gamma}LK\rightarrow\xi'n_{\gamma}'L'K')=\left(\frac{3R^{2}Ze}{4\pi}\right)^{2}\beta_{M}^{2}\times\nonumber\\
&&\left(B_{\xi,n_{\gamma}LK}^{\xi'n_{\gamma}'L'K'}G_{n_{\gamma}K}^{n_{\gamma}'K'}C_{KK'-KK'}^{L2L'}\right)^{2},
\label{E2prolate}
\end{eqnarray}
where $B$ and $G$ are denoted the contributions from the $\beta$ and $\gamma$ variables, while by $C$ the Clebsch-Gordon coefficient resulted from the matrix elements over the Euler angles. In the small angle approximation, $G$ takes the values $1$ for $\triangle K=0$, respectively $1/\sqrt{3q}$ for $|\triangle K=2|$.

Other relevant quantities for the present study are the probability density distribution of deformation in $\beta$,
\begin{equation}
\rho_{\xi,\nu}(\beta)=[F_{\xi,\nu}(\beta)]^{2}\beta^{4},
\label{density}
\end{equation}
the corresponding effective octic potential,
\begin{equation}
v_{eff}(\beta)=\frac{\Lambda+2}{\beta^{2}}+\beta^{2}+b_{1}\beta^{4}+b_{2}\beta^{6}+b_{3}\beta^{8},
\label{effpot}
\end{equation}
and the monopole transition $E0$ between the first excited $0_{2}^{+}$ state and the ground state $0_{1}^{+}$,
\begin{equation}
\rho^{2}(E0;0_{2}^{+}\rightarrow0_{1}^{+})=\left(\frac{3Z}{4\pi}\right)^{2}\beta_{M}^{4}\langle F_{2,0}(\beta)|\beta^{2}|F_{1,0}(\beta)\rangle^{2}.
\label{monopole}
\end{equation}

\section{Applications of the models to experimental data}

Three stable isobars are selected for the present study, $^{96}_{40}$Zr$_{56}$, $^{96}_{42}$Mo$_{54}$ and $^{96}_{44}$Ru$_{52}$, which are in vicinity of the harmonic oscillator proton shell closure $Z=40$, and the spin-orbit neutron shell closure $N=50$. This is a mass region where the former begins to fade away in favor of the spin-orbit shell structure. The competition between these two shell closures facilitates the shape coexistence phenomena through the so-called dual-shell mechanism \cite{Martinou1}. Previous studies made for these isobars \cite{Budaca4,Thomas,Cheng,Barbecho,Kumar,Kremer,Sazonov,Ramos} are in favor of this scenario. Consequently, new approaches are applied in the present work hoping to make further progress in elucidating this possibility.

\begin{figure}[t]
	\centering
	%\begin{tabular}{@{}cc@{}}
\resizebox{0.5\textwidth}{!}{
\includegraphics{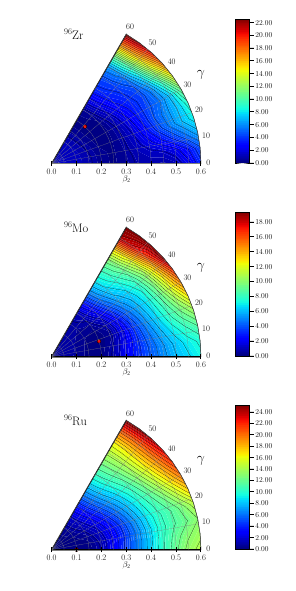}}
	%\end{tabular}
	\caption{(Color online) Potential energy surfaces in the $(\beta_{2},\gamma)$ plane for $^{96}$Zr, $^{96}$Mo and $^{96}$Ru obtained from the CDFT calculations with DD-PCX interaction. The red dot indicates the minimum point. The color scale, given in MeV units, shows the relative energy potential depth. }
	\label{fig1}
\end{figure}

Therefore, the ground state deformation for each of the three isobars is extracted from the minima of the potential energy surfaces shown in Fig. \ref{fig1}, which are determined with the CDFT using a DD-PCX parametrization.
The coordinates of the minima are given in Table \ref{tab2} and compared with the corresponding average deformation calculated with BMH-OP, respectively with the experimental values taken from the database \cite{ENSDF}.
\begin{table}[h!]
\caption{The $(\beta_{2},\gamma)$ ground state deformations, calculated with CDFT with DD-PCX parametrization and with BMH-OP for the isobars $^{96}$Zr, $^{96}$Mo and $^{96}$Ru, are presented together with the corresponding experimental $\beta_{2}$ values taken from the ENSDF data base \cite{ENSDF}.}
\begin{center}
\resizebox{6.cm}{0.9cm} {
\begin{tabular}{c|ccc}
\hline
\hline
$(\beta_{2},\gamma)$&$^{96}$Zr&$^{96}$Mo&$^{96}$Ru\\
\hline
\hline
CDFT             &$(0.200,48^{o})$&$(0.200,18^{o})$&$(0.100,0^{o})$\\
BMH-OP           &$(0.086,-)$     &$(0.163,-)$     &$(0.148,-)$\\
ENSDF            &$(0.060,-)$     &$(0.172,-)$     &$(0.155,-)$\\
\hline
\end{tabular}}
\end{center}
\label{tab2}
\end{table}
Thus, according to these results, one has an approximately oblate shape for $^{96}$Zr, a triaxial one for $^{96}$Mo, respectively a prolate one for $^{96}$Ru. Starting from these data, but also taking into account other properties which will be discussed in the following subsections, the excited states of the $^{96}$Zr isobar is finally described here within BMH-OP by considering an axially symmetric deformation, while those of the $^{96}$Mo and $^{96}$Ru involving a $\gamma$-unstable symmetry. The parameters of BMH-OP, fitted for the experimental data \cite{Abriola,Zielinska} of $^{96}$Zr, $^{96}$Mo and $^{96}$Ru are given in Table \ref{tab3} and they have been determined in the following way.
\begin{table}[h!]
\caption{Parameters of the Bohr-Mottelson Hamiltonian with Octic Potential (BMH-OP) fitted through the root mean square procedure (rms) for the experimental data \cite{Abriola,Zielinska} of the isobars $^{96}$Zr, $^{96}$Mo and $^{96}$Ru. The fitted parameters $b_{1}$, $b_{2}$ and $b_{3}$ lead to the following values of $\beta_w$: 3.96, 4.2, 2.6.  }
\begin{center}
\resizebox{8.5cm}{0.9cm} {
\begin{tabular}{cccccccc}
\hline
BMH-OP&\multicolumn{4}{c}{Free parameters}&&\multicolumn{2}{c}{Scaling factors}\\
\cline{2-5}\cline{7-8}
Isobar&$b_{1}$&$b_{2}$&$b_{3}$&$c$&$rms$&$\frac{\hbar^{2}}{2B}$ [keV]&$\beta_{M}$\\
\cline{2-5}\cline{7-8}
$^{96}$Zr&1.376&-0.263&0.012&$-$&0.050&1054&0.075\\
$^{96}$Mo&1.237&-0.228&0.010&$\;$0.070&0.108&198&0.144\\
$^{96}$Ru&10.56&-5.780&0.773&-0.062&0.047&471&0.103\\
\hline
\end{tabular}}
\end{center}
\label{tab3}
\end{table}
The free parameters $b_{1}$, $b_{2}$, $b_{3}$ and $c$ are fitted through the root mean square procedure (rms) for the experimental energies normalized to the energy of the first excited state of the ground band. The parameter $c$ intervenes only in the $\gamma$-unstable case to break the degeneracy of the energies over $\tau$. In turn, the corresponding wave functions remain independent on the parameter $c$. With other words, this parameter does not influence the electromagnetic transitions, for example. The $\beta_{w}$ limit, which defines the infinite square well potential, is constrained to intersect the tail of the octic potential. Thus, $\beta_{w}$ depends on the free parameters $b_{1}$, $b_{2}$ and $b_{3}$ and is calculated for each of the three isobars in caption of Table \ref{tab3}. Further, the scaling factor $\hbar^{2}/2B$ is fixed to exactly reproduce the experimental energy of the first excited state of the ground band. In this way, the correspondence in keV units between the theoretical energies and experimental ones is restored. For the $E2$ and $E0$ electromagnetic transitions, only the scaling factor $\beta_{M}$ remains to be determined since the other free parameters are already found. Thus, $\beta_{M}$ is fixed such that the $B(E2)$ transition between the first excited state of the ground band and the ground state to be exactly reproduced. Then, $\beta_{M}$ is used to calculate also the monopole $E0$ transition given by Eq. (\ref{monopole}).
Firstly, with the values of the parameters, the average values of the $\beta_{2}$ deformation have been calculated with BMH-OP, those shown in Table \ref{tab2}. To a first view, one can conclude that these values are very close to the experimental ones \cite{ENSDF} and somewhat closer to those provided by CDFT with DD-PCX for $^{96}$Mo and $^{96}$Ru. Concerning the $\gamma$ deformation within BMH-OP, this will be discussed in the next subsections in relation to the energies, electromagnetic transitions, energy potentials and probability density distribution of deformation. A correlation between all these quantities is made so that to extract information about the shape of each state, and consequently about the presence of the shape coexistence.

\subsection{The $^{96}$Zr isobar}
According to the results shown in Fig. \ref{fig1}, obtained with CDFT using DD-PCX, one has an approximately oblate shape  for the ground state of $^{96}$Zr: $(\beta_{2},\gamma)=(0.200,48^{o})$. Therefore, an axially-symmetric deformation would be appropriate for BMH-OP to describe the excited states of this isobar. Since the solution of the BMH-OP for a prolate deformation is already proposed \cite{Budaca2} and introduced in Section 2, one takes the advantage of the periodicity of the wave function in relation to $\gamma=n\pi/3$ ($n\in Z)$ such that to apply it for the oblate case. For example, as it is discussed in \cite{Fortunato}, $n=1$ leads to $\gamma=\pi/3$ (oblate shape) and to a projection of the total angular momentum on the $x$-axis as a good quantum number. On the other hand, if one skip this case and considers $n=3$,  a new oblate shape is obtained for $\gamma=\pi$, but this time with the projection of the total angular momentum on the $z$-axis. Thus, the formula for the energy in this case, for a zero-order approximation of the rotational term of the Hamiltonian, remains unchanged compared with the prolate case ($n=0$).
\begin{figure*}[t]
	\centering
	%\begin{tabular}{@{}cc@{}}
\resizebox{0.7\textwidth}{!}{
\includegraphics{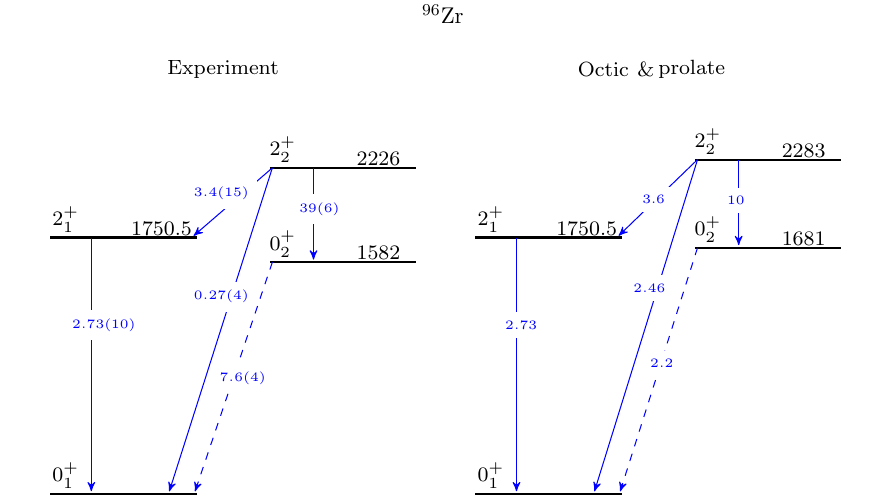}}
	\caption{(Color online) The experimental data for the lowest states of the $^{96}$Zr isobar \cite{Sazonov,Abriola,Zielinska} are compared with the calculated ones using the Octic $\&$ prolate approach. The energies and the $B(E2)$s are given in keV and W.u., respectively, while the monopole $E0$ transition (dashed arrow) is multiplied by a factor of $10^{3}$.}
	\label{fig2}
\end{figure*}
Concerning the quadrupole electromagnetic transitions, these are also not affected even if the $\gamma$ variable becomes  $\gamma-\pi$. According to Eq. (\ref{E2prolate}), the contribution from the functions in the $\gamma$ variable in this case is a constant $G$ which for the states of the ground and $\beta$ bands considered here is $1$.
The results obtained by applying this method are shown in Fig. 2. For this isobar there is no certain associated experimental $4_{1}^{+}$ state or higher states in the ground band. Also, the data for the $\gamma$ band is missing. More states can be found instead in the $\beta$ band, but only two of them fit better the present description. This can be an indication that other degree of freedom should be taken into account for the highest states of the $\beta$ band. Resuming our discussion to the considered states, one can remark that there is an overall good agreement between theory and experiment. With the fitted parameters, the effective potentials and probability density distributions of deformation are plotted in Fig. \ref{fig3}.
\begin{figure}
	\centering
	%\begin{tabular}{@{}cc@{}}
\resizebox{0.45\textwidth}{!}{
\includegraphics{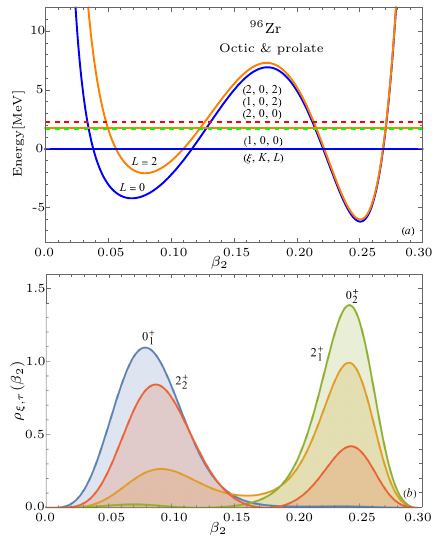}}
	\caption{(Color online) The effective potentials and the energy levels (panel (a)), respectively the probability density distributions of the $\beta_{2}$ deformation  (panel (b)), for the $0_{1}^{+}$, $2_{1}^{+}$, $0_{2}^{+}$, $2_{2}^{+}$ states of the $^{96}$Zr nucleus. }
	\label{fig3}
\end{figure}
The effective potential has two minima, a near-spherical and a well-deformed one, respectively. The second minimum is closer to $\beta_{2}=0.200$ calculated with the DD=PCX, which corresponds here to the single peak of the $0_{2}^{+}$ state, while the ground $0_{1}^{+}$ state is positioned above the less-deformed minimum. This behavior represents a clear indication of a presence of the shape coexistence, especially that usually the $0_{2}^{+}$ state has two peaks, its wave function having one node associated to the $\beta$ vibration. In turn, the $\beta$ vibration is almost frozen for $0_{2}$ due to the high barrier (maximum) of the effective potential, which separates the two minima and implicitly the two states. This is the reason for which the monopole $E0$ transition between these two states, presented in Fig. \ref{fig2} by a dashed arrow, is so small. There is no mixing of shapes for the two states, which is responsible for large monopole values. Only a negligible tail of the peak of the $0_{2}^{+}$ state is present above the less-deformed minimum. On the other hand, for the two $2^{+}$ states, which are higher in energy and have an effective potential with a slightly high near-spherical minimum, one has plots of the probability density distribution of deformation with two peaks like in the mirror. A reduced $\beta$ vibration is observed this time for the $2_{2}^{+}$ state, while the presence of the two peaks for the $2_{1}^{+}$ is related to the presence of the shape coexistence with mixing \cite{Budaca3} since one has no $\beta$ vibration node for the states of the ground band. Another important remark is that the deformation is reversed, each of the two states having a clear dominant peak and in opposite direction compared with the lower states in band. The results are similar with those presented in \cite{Sazonov}, where a different approach is used for the same states, respectively the corresponding wave functions are plotted instead of the square of the wave functions multiplied by the volume element. Therefore, the present study comes to strengthen this picture for the lowest states of the $^{96}$Zr characterized by the presence of the shape coexistence and mixing phenomena.

\subsection{The $^{96}$Mo isobar}
\begin{figure*}
	\centering
	%\begin{tabular}{@{}cc@{}}
\resizebox{1.02\textwidth}{!}{
\includegraphics{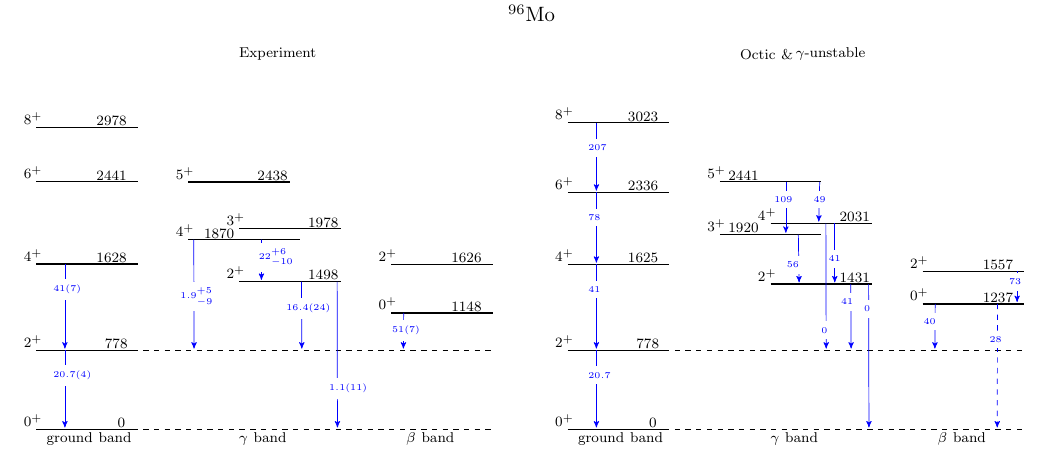}}
	\caption{(Color online) The experimental data for the lowest states of the $^{96}$Mo isobar \cite{Abriola} are compared with the calculated ones using the Octic $\&$ $\gamma$-unstable approach. The energies and the $B(E2)$s are given in keV and W.u., respectively, while the monopole $E0$ transition (dashed arrow) is multiplied by a factor of $10^{3}$.}
	\label{fig4}
\end{figure*}

According to the potential energy surface determined with CDFT using DD-PCX, shown in Fig. \ref{fig1}, the $^{96}$Mo isobar has a  triaxial deformation, with a considerable width for the potential well.
This makes it an appropriate candidate for the BMH-OP for $\gamma$-unstable symmetry. Previously, $^{96}$Mo was investigated with the Bohr-Mottelson Hamiltonian with Sextic Potential (BMH-SP) for $\gamma$-unstable symmetry getting a good agreement with the corresponding experimental data and evidencing fingerprints of the presence of the shape coexistence. The octic potential, being more general than the sextic potential, is expected to better encompass this phenomenon. Thus, the results obtained with BMH-OP for the lowest states of the ground, $\beta$ and $\gamma$ bands of the $^{96}$Mo isobar are given in Fig. \ref{fig4}.
\begin{figure*}
	\centering
	%\begin{tabular}{@{}cc@{}}
\resizebox{0.9\textwidth}{!}{
\includegraphics{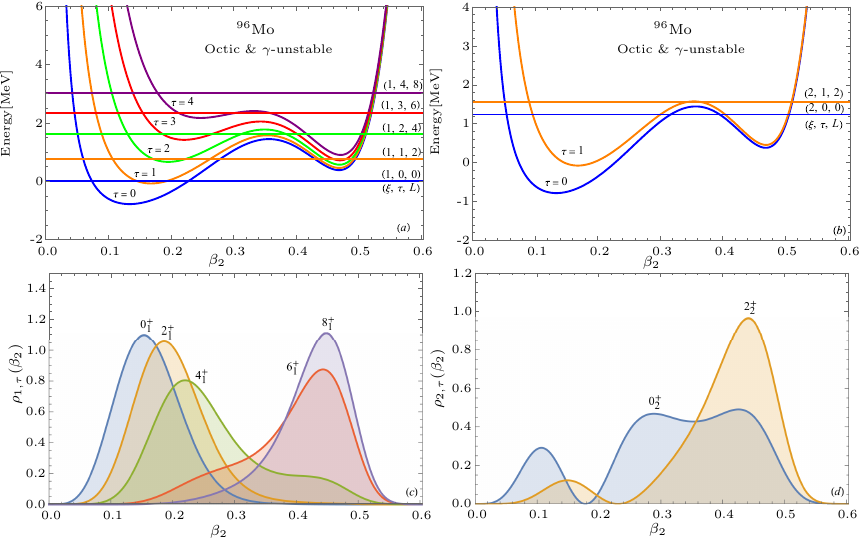}}
	\caption{(Color online) The effective potentials and the energy levels (panels (a) and (b)), respectively the probability density distributions of the $\beta_{2}$ deformation  (panels (c) and (d)), for the $0_{1}^{+}$, $2_{1}^{+}$, $4_{1}^{+}$, $6_{1}^{+}$, $8_{1}^{+}$, $0_{2}^{+}$, $2_{2}^{+}$ states of the $^{96}$Mo nucleus. }
	\label{fig5}
\end{figure*}
The energy spectrum is very well reproduced for all considered states, excepting for the experimental $4^{+}$ state at 1870 keV which is lower in energy than the $3^{+}$ state. Also, there is a good agreement for the low energy of the first excited $0_{2}^{+}$ state relative to the head of the $\gamma$ band, respectively to the energy of the first excited $2_{1}^{+}$ state of the ground band. This latter aspect is important because, according to a signature proposed in \cite{Martinou2}, if the quantity $\triangle E=E_{0_{2}^{+}}-E_{2_{1}^{+}}<800$ keV, the nucleus belongs to an island of shape coexistence. Thus, for $^{96}$Mo one has $\triangle E=370$ keV in Exp. and 459 keV in Th. which supports this possibility. Indeed, the plot of the probability density distribution of deformation for the $0_{2}^{+}$ state, shown in Fig. \ref{fig5}, has three peaks instead of two suggesting a coexistence with mixing since the ground state has a single peak above the less-deformed minimum. The mixing is in agreement also with the calculated monopole transition of 28, represented in Fig. \ref{fig4} with a dashed arrow. This value corresponds to a moderate mixing which can be put in correspondence with the smaller peak of the $0_{2}^{+}$ state positioned above the less-deformed minimum.
Instead, the $2_{2}^{+}$ state has two peak with a dominant one above the well-deformed minimum, which reflects an attenuation of the $\beta$ vibration. An interesting behavior is observed in panels (a) and (c), where the increasing in energy of the less-deformed minimum of the potential describes a shape transition in band. The $0_{1}^{+}$ and $2_{1}^{+}$ states have a single peak centered above the less-deformed minimum, which is understandable given the fact that the energy of the ground state intersect the potential only above the first minimum, while the $2_{1}^{+}$ state only slightly the second minimum. Instead, according to panel (a) of Fig. \ref{fig5}, the effective potential for $\tau=2$ has two minima almost degenerated with the energy of the $4_{1}^{+}$ state slightly below the barrier. The corresponding plot of the probability density distribution of deformation has a dominant peak above the less-deformed minimum and a smaller flat peak above the well-deformation. Something similar is observed for the $6_{1}^{+}$ state even if for these two states one has no-node. The shape transition is completed by the $8_{1}^{+}$ state which has a single peak around the well-deformed minimum of the potential. This change in shape in the ground band, respectively the shape coexistence with mixing, influence also the values of the $B(E2)$ transitions. For example, a contraction in strength is observed in theory for the $B(E2;4_{1}^{+}\rightarrow2_{1}^{+})$ transition of which value of $41$ W.u. matches perfectly the experimental data of $41(7)$ W.u.. This contraction is related to the fact that the $4_{1}^{+}$ state is in the "critical" point of the shape transition having the dominant peak lower than that of the $2_{1}^{+}$ state. In turn, the $B(E2;8_{g}^{+}\rightarrow6_{g}^{+})$ transition is much larger than the previous one. The $6_{1}^{+}$ state has the dominant peak above the well-deformed minimum, as it is the case for the $8_{1}^{+}$ state, but with the peak lower than that for $8_{1}^{+}$. An unusually large value is measured for the $B(E2;0_{2}^{+}\rightarrow2_{1}^{+})$ transition of $51(7)$ W.u.. This is even larger than the $B(E2;4_{1}^{+}\rightarrow2_{1}^{+})$ transition. In theory one has $40$ W.u. for this transition, which is close to the experimental value if the experimental errors are taken into account. An explanation for this large transition is related to the presence of the shape coexistence with mixing given by the three peaks of the $0_{2}^{+}$ state, two of them highly overlapping with the single peak of the $2_{1}^{+}$ state. Concerning the $B(E2)$s related to the $\gamma$ band, the experimental $B(E2;2_{\gamma}^{+}\rightarrow2_{1}^{+})$ is of $16.4(24)$W.u. in experiment and more than double in theory. Actually, due to the $\gamma$-unstable symmetry of the Hamiltonian, the $2_{\gamma}^{+}$ and $4_{1}^{+}$ states are degenerated and consequently their corresponding $B(E2)$ transition to the $2_{1}^{+}$ state is the same. Another aspect is related to the selection rules, given by Eq. (\ref{tauop}), which forbid some $B(E2)$s such as the $B(E2;4_{\gamma}^{+}\rightarrow2_{1}^{+})$ and $B(E2;2_{\gamma}^{+}\rightarrow0_{1}^{+})$ transitions. These two transitions are indeed very small in experiment, but not enough small so that to be considered negligible. Also, this experimental $4^{+}$ state at 1870 keV is incorporated in the $\gamma$ band even if it is lower in energy than the $3_{\gamma}^{+}$ state. This consideration has been motivated by the measured $B(E2)s$ from this state to the $2_{1}^{+}$ state, already discussed, respectively to the $2_{\gamma}^{+}$ state. The latter $B(E2)$ transition is of $22_{-10}^{+6}$ W.u. in experiment compared with 41 W.u. in theory. These two $B(E2)$ transitions are not perfectly reproduced in theory, but still demonstrates a consistent description.

\begin{figure*}
	\centering
	%\begin{tabular}{@{}cc@{}}
\resizebox{1.01\textwidth}{!}{
\includegraphics{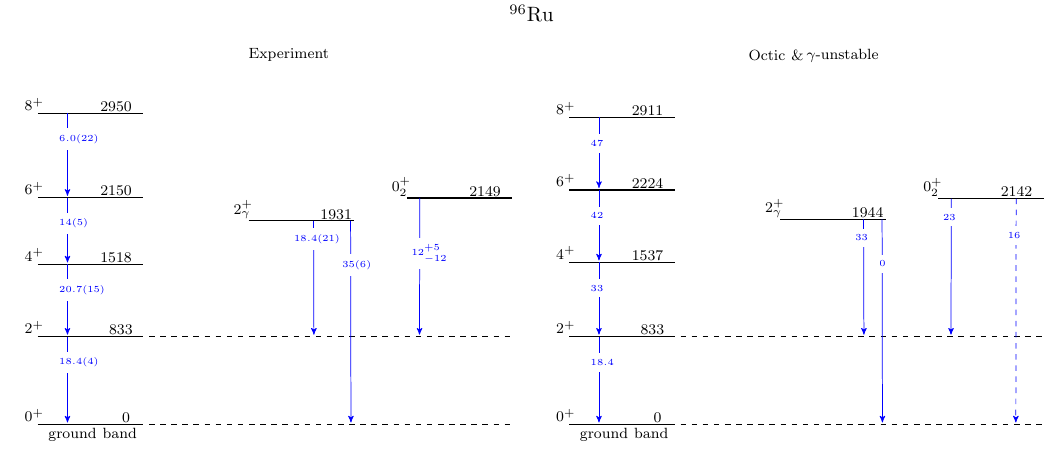}}
	\caption{(Color online) The experimental data for the lowest states of the $^{96}$Ru isobar \cite{Abriola} are compared with the calculated ones using the Octic $\&$ $\gamma$-unstable approach. The energies and the $B(E2)$s are given in keV and W.u., respectively, while the monopole $E0$ transition (dashed arrow) is multiplied by factor of $10^{3}$. }
	\label{fig6}
\end{figure*}

\begin{figure}
	\centering
	%\begin{tabular}{@{}cc@{}}
\resizebox{0.45\textwidth}{!}{
\includegraphics{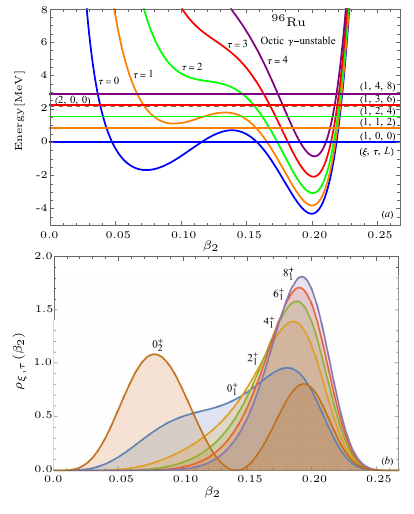}}\\
	\caption{(Color online) The effective potentials and the energy levels (panel (a)), respectively the probability density distributions of the $\beta_{2}$ deformation  (panel (b)), for the $0_{1}^{+}$, $2_{1}^{+}$, $4_{1}^{+}$, $6_{1}^{+}$, $8_{1}^{+}$, $0_{2}^{+}$ states of the $^{96}$Ru nucleus. }
	\label{fig7}
\end{figure}

\subsection{The $^{96}$Ru isobar}

For this isobar, the corresponding potential energy surface shown in Fig. \ref{fig1} indicates a minimum at $\beta_{2}=0.1$ and $\gamma=0^{o}$, which corresponds to a prolate shape judging by the axial deformation. Nevertheless, the $\beta_{2}$ deformation is quite small, being closer to a spherical limit. Also, the experimental energy ratio $R_{4/2}=1.82$ is closer more to the spherical vibrator for which this ratio is 2.
Moreover, the energies of the excited states of the ground band don't follow a prolate rotor structure evolution. Instead of having a rising rate proportional with $L(L+1)$, actually the energy contracts with the increasing of the spin. For example the difference in energy, starting from the $2_{1}^{+}$ state, is 833 keV, 685 keV, 632 keV and 801 keV. Concerning the $\beta$ and $\gamma$ bands, there is no certain states associated with them, but we considered in the present study the first excited $0_{2}^{+}$ state at 2149 keV, respectively the $2^{+}$ state at 1931 keV for the head of the $\gamma$ band. Initially, we applied the BMH-OP with prolate deformation, but without success in reproducing these states accordingly. This attempt confirmed somehow that the structure of this isobar is other than prolate or coexistence between near-spherical and prolate shapes. Instead, the BMH-OP with $\gamma$-unstable symmetry fitted very well this data as it can be seen from Fig. \ref{fig6}.
For example, the energy levels are almost perfectly reproduced and more important the contraction in energy in the ground band: 833 keV, 704 keV, 687 keV, 687 keV. Looking to the experimental $B(E2)$ values in the ground band, one can also remark a contraction trend in strength, which is seen somehow also in theory where the growth rate of the $B(E2)$ transition is slowing down with the increasing of the spin. Concerning the experimental $B(E2;0_{2}^{+}\rightarrow0_{1}^{+})$ and $B(E2;2_{\gamma}^{+}\rightarrow2_{1}^{+})$ transitions of $12_{-12}^{+5}$ W.u. and 18.4(21) W.u., respectively, they are relatively close to the corresponding theoretical values of 23 W.u. and 33 W.u., respectively. A large discrepancy, however, arises for the $B(E2;2_{\gamma}^{+}\rightarrow0_{1}^{+})$ transition which is forbidden in theory, while in experiment is very large compared with the rest of the $B(E2)$ transitions for this isotope, namely of 35(6) W.u..
It must be mentioned that this $B(E2)$ value falls out the systematic trend seen in the rest of the isotopic chain, namely $1.04^{+17}_{-14}$, $2.0(4)$, $1.14(15)$ and $2.8(5)$ W.u. for $^{98}$Ru \cite{Chen}, $^{100}$Ru \cite{Singh}, $^{102}$Ru \cite{Frenne1} and $^{104}$Ru \cite{Blachot}, respectively.
Thus, this large value of 35(6) W.u. doesn't match either the strength scale of the $B(E2)$ transition values for $^{96}$Ru, nor the values for the $B(E2;2_{\gamma}^{+}\rightarrow0_{1}^{+})$ transition of the next isotopes. Regarding the presence of the shape coexistence for this isobar, this information can be extracted by analyzing the plots from Fig. 7.
The effective potential has two clear minima for $\tau=1$ and $\tau=2$, a near-spherical one and a well-deformed one. The near-spherical minim is slowly lost for higher excited states. Because of that, the single peak of the excited states of the ground band becomes even more centered around the well-deformed minimum. Instead, the energy level of the ground state, being below the barrier separating the two minima, leads to an approximately two-peaks structure in the probability density distribution of deformation. On the other hand, the first excited $0_{2}^{+}$ state has two peaks like in the mirror compared with those of the ground state. This picture corresponds within the frame of the present model to a presence of the shape coexistence with mixing between an approximately spherical shape and a $\gamma$-unstable one.

\section{Conclusions}
Applications of the Covariant Density Functional Theory with a density-dependent point-coupling X interaction and of the Bohr-Mottelson Hamiltonian with octic potential (BMH-OP) for axially-symmetric and $\gamma$-unstable deformations to the lowest collective states of the $^{96}$Zr, $^{96}$Mo and $^{96}$Ru isobars shown that not only the shape coexistence and mixing phenomena are presented in these nuclei, but also that they lead to completely different structure of the states than usually. For example, a contraction in both energy and $B(E2)$ can manifest in band with the increasing of the spin, while a large strength of the $B(E2)$ transition between states of different bands can be observed if the shape coexistence with mixing is present.

Concretely, a coexistence with and without mixing, respectively, between an approximately spherical shape and an oblate one is proposed for the lowest states of the $^{96}$Zr isobar. Instead, for the $^{96}$Mo and $^{96}$Ru isobars, a coexistence with mixing between an approximately spherical shape and a $\gamma$-unstable one is found. Additionally, a complete shape transition is observed in the ground band of the $^{96}$Mo from an approximately spherical shape to a more deformed one crossing a point where the effective potential has two minima almost degenerated. Making the analogy with the ground-state shape phase transitions, we can call this phenomenon a dynamical shape transition since it is a function of the excitation energy.

Concluding, the present study offers a detailed analysis of the lowest states of these three isobars coming with a description which is consistent with the experimental observations.

\section*{Acknowledgments}
This work was supported by grants of the Ministry of Research, Innovation and Digitization, CNCS-UEFISCDI, project number PN-IV-P1-PCE-2023-0273, within PNCDI IV, and project number PN-23-21-01-01/2023, respectively through computational resources (www.marwan.ma) of HPC- MARWAN
 provided by the National Center for Scientific and Technical Research (CNRST) in Rabat, Morocco.

% BibTeX users please use
% \bibliographystyle{}
% \bibliography{}

\end{document}